\documentclass[runningheads]{llncs}
\usepackage[T1]{fontenc}
\usepackage{lmodern}
\usepackage{booktabs}
\usepackage[utf8]{inputenc}

\usepackage[inline]{enumitem}
\usepackage[normalem]{ulem}
\usepackage[skins,breakable]{tcolorbox}
\usepackage{amsmath,amsfonts}
\usepackage{array}
\usepackage{hyperref}

\usepackage{microtype}
\usepackage{framed}
\usepackage{lipsum}
\usepackage{listings}
\usepackage[capitalize]{cleveref}
\crefname{section}{Sect.}{sections}
\Crefname{section}{Section}{Sections}
\usepackage{longtable}
\usepackage{mathtools}
\usepackage{multirow}
\usepackage{pbox}
\usepackage{relsize}
\usepackage{rotating}
\usepackage{tabularx}
\usepackage{todonotes}
\usepackage{url}
\usepackage{subcaption}
\usepackage{graphicx}
\usepackage{xspace}
\usepackage{tikz}
\usetikzlibrary{shapes.geometric}
\usetikzlibrary{arrows, shapes}
\usepackage{tcolorbox}

\definecolor{xtextBlue}{RGB}{42,8,255}
\newcommand{\bt}{behavior tree\xspace}
\newcommand{\bts}{behavior trees\xspace}
\newcommand{\BTCPP}{\texttt{Behavior\-Tree.CPP}\xspace}
\newcommand{\pytrees}{\lstinline{py_trees}\xspace}

\usepackage{ifthen}
\usepackage{amssymb}

\newboolean{showcomments}
\setboolean{showcomments}{true} 
\ifthenelse{\boolean{showcomments}}
{\newcommand{\nb}[2]{
		\fcolorbox{gray}{yellow}{\bfseries\sffamily\scriptsize#1}
		{\sf\small$\blacktriangleright$\textit{#2}$\blacktriangleleft$}
	}
	
}
{\newcommand{\nb}[2]{}
	
}

\definecolor{blue(ncs)}{rgb}{0.0, 0.53, 0.74}
\AtBeginDocument{%
  }

\definecolor{backcolourX}{rgb}{1,1,1}
\definecolor{maroon}{rgb}{0.5,0,0}
\definecolor{blue-violet}{rgb}{0.54, 0.17, 0.89}

\definecolor{backcolour}{rgb}{0.95,0.95,0.92}
\definecolor{maroon}{rgb}{0.5,0,0}
\definecolor{blue-violet}{rgb}{0.54, 0.17, 0.89}
\definecolor{charcoal}{rgb}{0.21, 0.27, 0.31}

\lstdefinestyle{xtextstyle}{
	language=Java,
	basicstyle=\footnotesize ,
	columns=fullflexible,
	backgroundcolor=\color{backcolourX},
	keywordstyle=\color{violet},
	morekeywords={
		STRING, indent},
	stringstyle=\color{blue},
	classoffset=1,
	morekeywords={True},
	keywordstyle=\color{charcoal},
	morekeywords={
		BehaviorTree, TreeNode, LeafNode, ControlNode, DecoratorNode, SequenceNode, ParallelNode, RepeatNode, ConditionNode, ActionNodeBase, Quality, QualityRequirement,InverterNode, SubTree}
}

\lstdefinestyle{xmlstyle}{
	language=XML,
	columns=fullflexible,
	backgroundcolor=\color{backcolour},
	keywordstyle=\color{red},
	stringstyle=\color{blue-violet},
	morekeywords={
		main_tree_to_execute,name,num_attempts,_description,_successIf,_failureIf},
	classoffset=1,
	morekeywords={False},
	keywordstyle=\color{blue},
	morekeywords={
		root,BehaviorTree,SequenceStar,Sequence,Script,Retry, Fallback,ScriptCondition,Inverter,SubTree}
}

\begin{document}


\onecolumn 
\pagestyle{empty} 
\begin{center}
	\large\bfseries  
	This version of the contribution has been accepted for publication at REFSQ25 after peer review, but is not the Version of Record and does not reflect post-acceptance improvements, or any corrections. The Version of Record is available online at: https://doi.org/10.1007/978-3-031-88531-0\_24. Use of this Accepted Version is subject to the publisher’s Accepted Manuscript terms of use https://www.springernature.com/gp/open-science/policies/accepted-manuscript-terms".
\end{center}

\setcounter{page}{1} 

\title{Extending Behavior Trees for Robotic Missions with Quality Requirements}

\author{Razan Ghzouli\inst{1}\and
	Rebekka Wohlrab\inst{1,2} \and
	Jennifer Horkoff\inst{1}}
\authorrunning{R. Ghzouli et al.}
%
\institute{
Chalmers University of Technology and University of Gothenburg, Sweden \and Carnegie Mellon University, Pittsburgh, USA}

\maketitle

\begin{abstract}

\textbf{Context and motivation:} In recent years, behavior trees have gained growing interest within the robotics community as a specification and control-switching mechanism for the different tasks that form a robotics mission. 
\textbf{Problem:} Given the rising complexity and prevalence of robotic systems, it is increasingly challenging and important for practitioners to design high-quality missions that meet certain qualities, for instance, to consider potential failures or mitigate safety risks.
In software requirements engineering, quality or non-functional requirements have long been recognized as a key factor in system success. 
Currently, qualities are not represented in behavior-tree models, which capture a robotic mission, making it difficult to assess the extent to which different mission components comply with those qualities.
\textbf{Principal ideas/results:} In this paper, we propose an extension for behavior trees to have qualities and quality requirements explicitly represented in robotics missions. We provide a meta-model for the extension, develop a domain-specific language (DSL), and describe how we integrated our DSL in one of the most used languages in robotics for developing behavior trees, \BTCPP. A preliminary evaluation of the implemented DSL shows promising results for the feasibility of our approach and the need for similar DSLs.
\textbf{Contribution:} Our approach paves the way for incorporating qualities into the behavior model of robotics missions. This promotes early expression of qualities in robotics missions, and a better overview of missions' components and their contribution to the satisfaction of quality concerns.

\vspace{-0.2cm}
\end{abstract}

%



\vspace{-0.5cm}
\section{Introduction}
\vspace{-0.3cm}
\label{sec:intro}

In recent years, behavior trees have emerged as the preferred behavior model for defining missions and coordinating task-switching in cyber-physical robots, especially when reactiveness is crucial.
Behavior trees gained  popularity in the robotic community for their modularity, expressiveness, readability, maintainability and flexibility \cite{iovino2023programming,colledanchise2018behavior}
Using behavior trees can help practitioners when designing, understanding, and modifying robotic missions. 
However, the current model of behavior trees only captures the tasks and actions of a mission \cite{ghzouli2023behavior}.
At the same time, there are many other concerns that stakeholders can have when working with robotic missions \cite{garcia2023software}.
Many of these concerns are connected to qualities, such as safety, security, and performance. 
Riley et al.'s guidelines for enhancing human-robot interaction (HRI) \cite{riley2010situation} emphasized the importance of going beyond basic data about the robotic missions and integrating information, such as mission-relevant requirements, that facilitate better situational awareness.
When working with current behavior trees, qualities are often not explicitly represented or considered~\cite{iovino2022survey,ghzouli2023behavior}.

If qualities are not considered during robotic mission design- and run-time, there is a risk that they come as an afterthought and are not effectively supported.
In early design time, it is beneficial to indicate the important qualities, so that they can be systematically elicited, even if concrete measurements are not yet known~\cite{yu1997towards}.
During late design time,  quality requirements should be specified more concretely, so that they can be monitored and measured at run-time~\cite{vierhauser2016requirements}.
During run-time, stakeholders might want to follow up on the satisfaction of those requirements.
To the best of our knowledge, there exists no quality-focused model for behavior trees that supports these activities.

The robotic domain suffers from a lack of software engineering practices. For example, the robotic 2020 multi-annual report ICT-2017 has stressed the importance of adopting model-driven (MD) methods to reduce the complexity of robotics systems and improve the maintainability of systems \cite{robotics2017robotics}. Domain-specific languages (DSLs), an MD approach, are becoming popular in robotics due to their expressiveness,  ease of use, and ability to promote communication between developers and domain experts \cite{casalaro2022model,nordmann2014survey}. 
Currently, available behavior-tree DSLs do not support capturing and monitoring robotics mission requirements \cite{ghzouli2023behavior,iovino2022survey}.

In this paper, we propose an extension for behavior trees to explicitly represent quality concerns and quality requirements in robotics missions.
Our approach supports the explicit representation of quality concerns (qualities and quality requirements) by providing a foundational model and creating a fit-for-purpose DSL that the robotics community can use.
We provide a meta-model for behavior trees including qualities and quality requirements.
By providing a DSL as a concrete representation of the meta-model, we can support developers already at design
time to specify quality concerns of the robotic missions. We use MD engineering to create our DSL, and automatically convert our DSL code into a widely-used robotics behavior-tree framework, \BTCPP \cite{faconti2019mood2be}. By offering the auto-generation to \BTCPP code, we promote the integration of our work with existing frameworks in robotics, and we integrate the quality concerns into \BTCPP.
We illustrate and demonstrate the practical applicability
of our  approach by applying the DSL to an open-source project in
the context of a laboratory-robot mission.
We further assess the feasibility of the proposed DSL by conducting a preliminary evaluation with six practitioners and researchers.  All presented materials in this paper are available in our online appendix \cite{appendix:online}.


\vspace{-0.5cm}
\section{Background}
\vspace{-0.3cm}
\label{sec:background}

Qualities, also referred to as non-functional requirements or quality attributes, have long been a focus in requirements engineering research and practice~\cite{chung1999non,yu1997towards,eckhardt2016non}.  System qualities have been considered as part of systems and software engineering standards such as ISO 25010~\cite{ISO25010}. The general message has been that such system qualities often go neglected in practice, as they can be more difficult to measure and define than their functional counterpart.  However, ignoring system qualities such as performance, safety, usability, and security is disastrous for the success of a system.  We posit that the same message holds for robotic systems.

Previous work in quality requirements has introduced the notion of `early' and `late' requirements engineering~\cite{yu1997towards}.  In the early stages, it is important to identify all relevant qualities for the system, but not yet force stakeholders to quantify such requirements. In such stages, we speak of `satisficing' (sufficiently satisfying) requirements, as per Simon~\cite{simon1956rational}. In later stages of requirements engineering, such qualities should be quantified and measurable, and thus satisfiable in the full sense. 
We use these ideas as inspiration in this work to capture qualities at two stages of the robotics mission lifecycle.

\begin{table}[t]
		\vspace{-5mm}
        \footnotesize
	\caption{The visual syntax of Node types in behavior trees.}
        \centering
    \scalebox{0.8}{
        	\begin{tabularx}{\linewidth}{XX||XX}
		
		\textbf{Node type} &\textbf{Symbol} & \textbf{Node type} &\textbf{Symbol}\\
		\midrule
		\lstinline!Sequence! & \begin{tikzpicture} \node [ draw, rectangle, minimum width=0.5cm, minimum height=0.25cm] at(0,0){$\rightarrow$}; \end{tikzpicture} &	\lstinline!Decorator! & \begin{tikzpicture} \node [draw, diamond, minimum width=0.5cm, minimum height=0.14cm] at(0,0){}; \end{tikzpicture}
		\\ 
		\lstinline!Selector! & \begin{tikzpicture} \node [ draw, rectangle, minimum width=0.5cm, minimum height=0.25cm] at(0,0){?}; \end{tikzpicture}  &
		\lstinline!Action! & \tikz \fill [gray]  rectangle (0.5cm,0.25cm);

		\\ 
		\lstinline!Parallel! & \begin{tikzpicture} \node [ draw, rectangle, minimum width=0.5cm, minimum height=0.25cm] at(0,0){ $\rightrightarrows$}; \end{tikzpicture} &
		\lstinline!Condition! & \tikz \draw (0cm,0cm) ellipse[x radius=0.25cm,y radius=0.15cm]; 

		\\ 
		
		\bottomrule

		\label{tab:btsyntax}	
	\end{tabularx}
	
    }
    \vspace{-10mm}
\end{table}

\textbf{Behavior Trees:} Behavior tree is a behavior model for coordinating the control-switching between various tasks involved in executing a mission.  
Behavior trees are seen as graphical models shaped as directed trees. They consist of a root node, non-leaf nodes called control-flow nodes and leaf nodes called execution nodes. The execution is managed by a tick, which is a signal issued according to a specific frequency \cite{colledanchise2018behavior}. The tick traverses the tree from the root down to its children according to the semantics of the control-flow nodes. A node is executed only upon receiving ticks. A ticked node returns to its parent one of the three statuses: success upon achieving its goal, failure if unsuccessful, or running if its execution is ongoing. 

The basic types of control-flow nodes are sequence, selector, decorator and parallel. Sequence nodes require all their children to succeed for the sequence node themselves to succeed, while selector nodes only need one child to succeed for their success status. Decorator nodes enable complex control flow, such as for and while loops. The extensibility of the decorators resulted in multiple off-the-shelf types provided by behavior-tree languages \cite{ghzouli2023behavior}. A decorator node only has one child, while sequence and selector nodes allow composing multiple children, hence classified as composite nodes.
Execution nodes are classified into action nodes (e.g., moving the robot to a specific location) and condition nodes (e.g., checking if a robot's battery is lower than a specific threshold). \Cref{tab:btsyntax} presents the visual syntax of node types in behavior trees \cite{ghzouli2023behavior,iovino2022survey,colledanchise2018behavior}. 

When working with cyber-physical robots that use the Robot Operating System (ROS), \BTCPP (\url{https://www.behaviortree.dev/}) and \pytrees (\url{https://py-trees.readthedocs.io/}) are popular behavior-tree DSLs \cite{ghzouli2023behavior,iovino2022survey}. \pytrees is the main behavior-tree DSL in the Python community, and \BTCPP is the leading DSL in C++.
The language design of \BTCPP is influenced by its origins within the European project RobMoSys project, specifically the MOOD2Be \cite{faconti2019mood2be}, which emphasized model-driven approaches and aimed to improve the reusability of robotic software components. \BTCPP adoption as a core component of the ROS navigation stack, Navigation 2, highlights its prominence within the robotics community compared to \pytrees \cite{macenski2020marathon}.
\BTCPP has an available graphical-user interface, called Groot, (\url{https://github.com/BehaviorTree/Groot}).
Groot can be used during design time as an editor and during execution time to monitor the running \bts. There are three ways to create \bts in \BTCPP. The first is by writing the \bts directly in C++ code.
The second involves using the Groot GUI to construct the \bts visually through a drag-and-drop interface. This process generates an XML-like file representing the tree syntax. The third, \bts can be defined directly in static files using the \BTCPP  XML-like language. The developers need to write the functionality behind each execution node in C++, e.g., write the code for pick, while for control nodes their implementations (code) are provided by \BTCPP. 

\begin{figure*}[h]
		\vspace{-0.5cm}
	\begin{center}
	\includegraphics[
		width=0.9\linewidth
		]{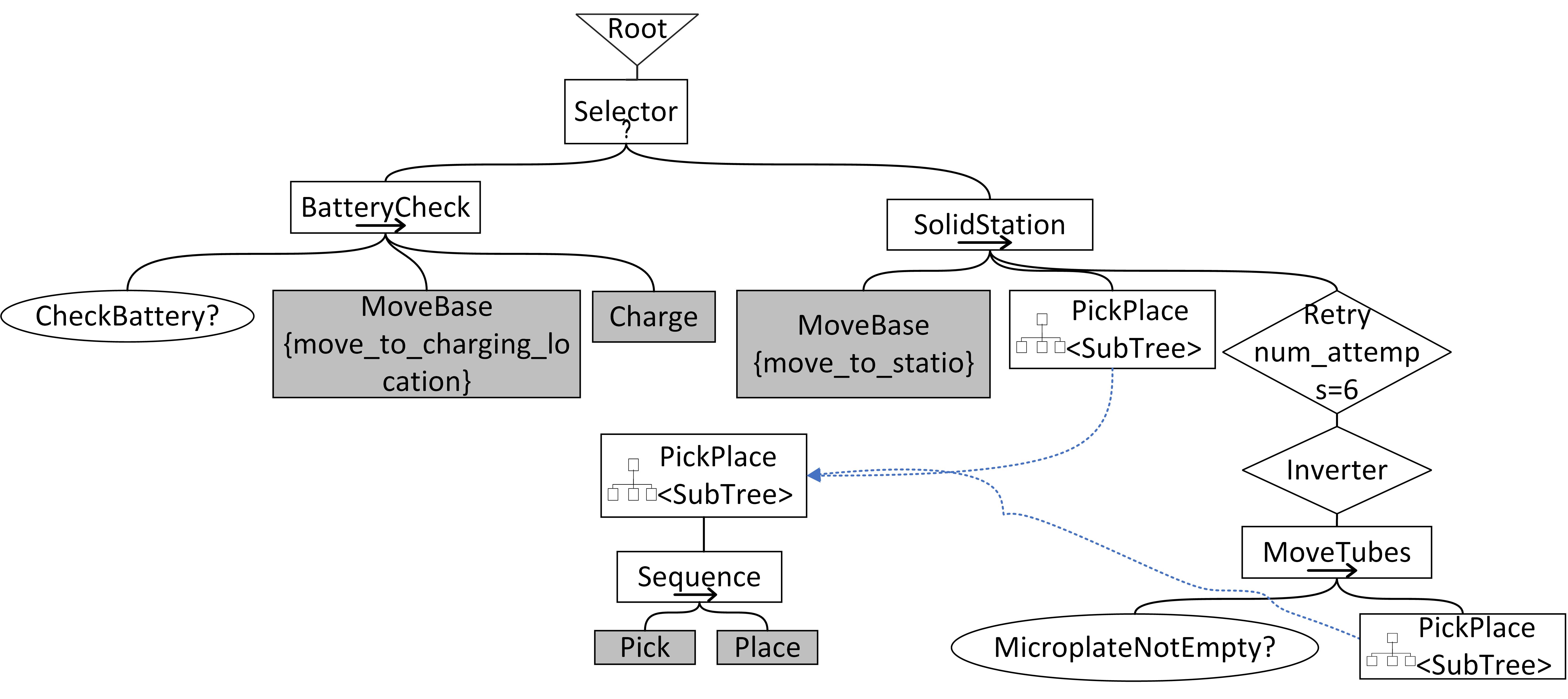}
	\end{center}
		\vspace{-5mm}
	\footnotesize\caption[Behavior Tree Example]{A \bt of the mobile laboratory robot. The dotted arrows show \lstinline!PickPlace! sub-tree expanded.}
	\label{fig:figure-1}
	
		\vspace{-6mm}
	
\end{figure*}

\textbf{Illustrative Example:} We present a behavior tree example of a mobile laboratory robot inspired by Burger et al.~\cite{burger2020mobile}'s mobile robotic chemist.
\Cref{fig:figure-1} shows an example of a behavior-tree model for the mobile laboratory robot. The main goal of the mission is to move test tubes to an automated sample-handling system. The mission has two tasks: moving the tubes \lstinline!SolidStation! and charging the robot when needed \lstinline!BatteryCheck!, each of them having multiple skills (a.k.a. actions) involved to achieve them.
The main operation is placed in the sub-tree \lstinline!SolidStation!: it consists of the robot moving to the designated station \lstinline!MoveBase!, picking the micro-plate and placing it near the automated sample-handling machine represented in the \lstinline!PickPlace! sub-tree, and moving the tubes out of the micro-plate \lstinline!MoveTubes! sequence. 
If the microplate is empty, the first node in the \lstinline!MoveTubes! sequence fails, which is converted into a success by an \lstinline!Inverter! such that the sequence of moving tubes is complete. Otherwise, the robot keeps repeating the same operation (checking if the microplate is not empty, then picking and placing sequence) up to six times, as long as the microplate is not empty. The \lstinline!MoveTubes! sequence is placed in a loop until all children succeed. The mission accounts for low-battery through the \lstinline!BatteryCheck! sequence: checking if the battery is lower than a specific percentage, moving to the charging station, and charging. The \lstinline!BatteryCheck! sequence has priority over the \lstinline!SolidStation!, meaning that if at any point the \lstinline!BatteryLow?! condition node is true, the current active node is interrupted and the robot executes the \lstinline!BatteryCheck! sequence.

In this illustration, the performance of the robot in terms of time and other factors is key.  If the robot operates too slowly, the automation benefits provided by robotics are not worthwhile.  Likewise, given the security-sensitive nature of laboratories (e.g., patients' data on the tubes), the security of the robotic process is also important to consider at all stages of design. Examples of important qualities for this mission are provided in \cref{subsec:concretemodel}.

\begin{figure*}[t]
	\begin{center}
		\includegraphics[width=0.7\linewidth]{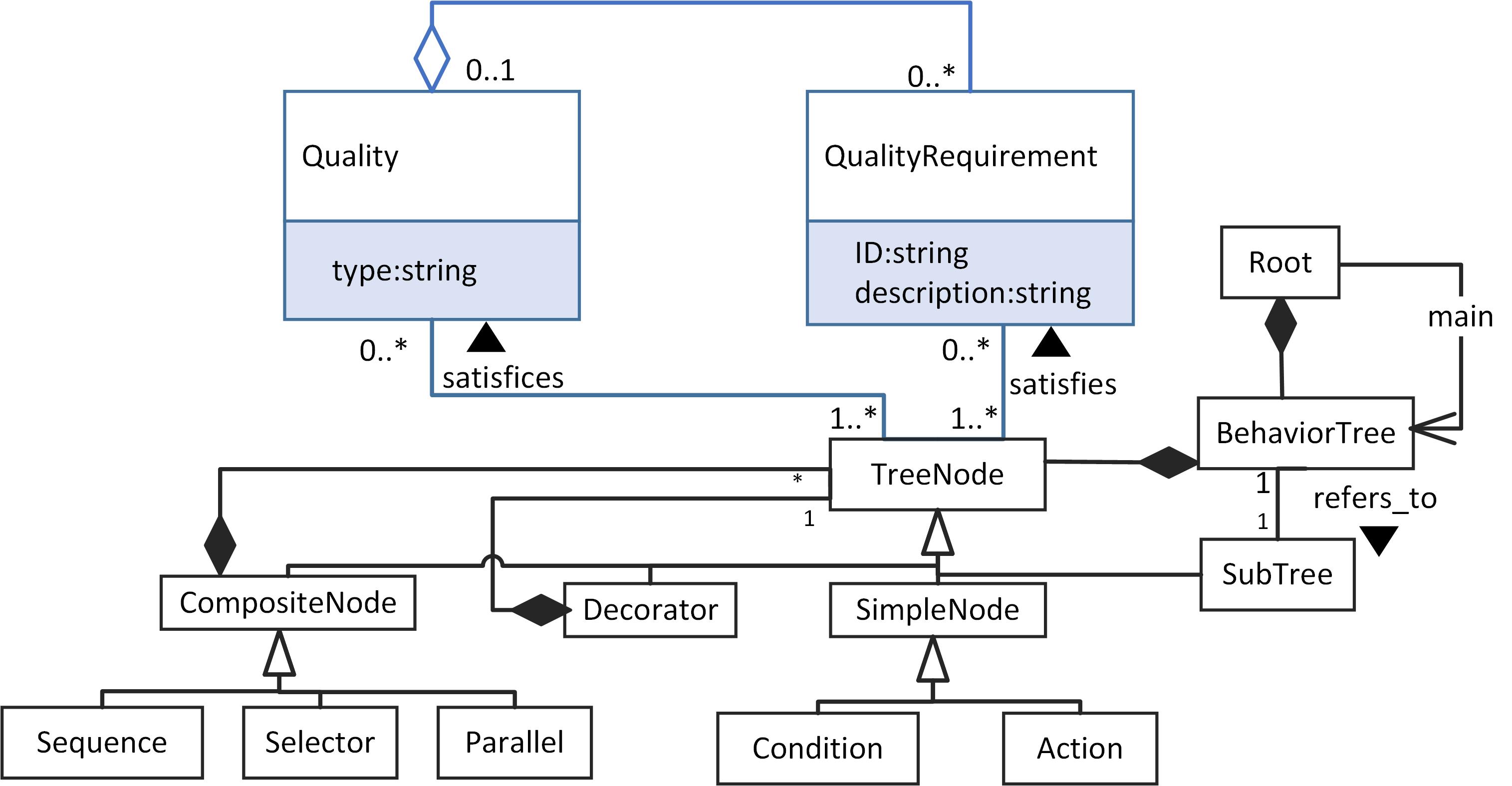}
	\end{center}
		\vspace{-5mm}
	\footnotesize\caption[Proposed meta-model]{Our quality-focused meta-model extension of the behavior-tree model}%
	\label{fig:figure-2}
	
	\vspace{-6mm}
	
\end{figure*}

\vspace{-0.3cm}
\section{Proposed Approach}
\vspace{-0.2cm}
\label{sec:approach}

Our goal is to explicitly
represent qualities and quality requirements in robotics missions.
We propose an extension of behavior trees to capture relevant quality concerns throughout the lifecycle of a mission. 
In our case, we consider the lifecycle stages of early and later design time, as well as the implementation (coding) of the mission. In the following, we describe the meta-model for extending the behavior-tree model with quality specifications.

\vspace{-0.3cm}
\subsection{Behavior-Tree Meta-Model with Quality Concerns}
\vspace{-0.2cm}
\label{subsec:metamodel}
\Cref{fig:figure-2} represents a meta-model that describes our proposed extension for the behavior-tree meta-model that considers the qualities and quality requirements of the robotics missions. The representation of the meta-model is divided into black parts (lower part of the figure) and blue parts (upper part of the figure). In black is the behavior-tree meta-model that was adopted from Ghzouli et al. \cite{ghzouli2023behavior}. Although this meta-model was reverse-engineered for a specific behavior-tree language, it is generalizable as a meta-model for behavior trees since it represents the basic components of behavior trees without going into the specification of the represented language. 
The new classes are in blue. These classes extend the behavior-tree meta-model to handle qualities and quality requirements. 

\begin{figure*}[t]
	\begin{center}
		\includegraphics[
		width=0.8\linewidth
		]{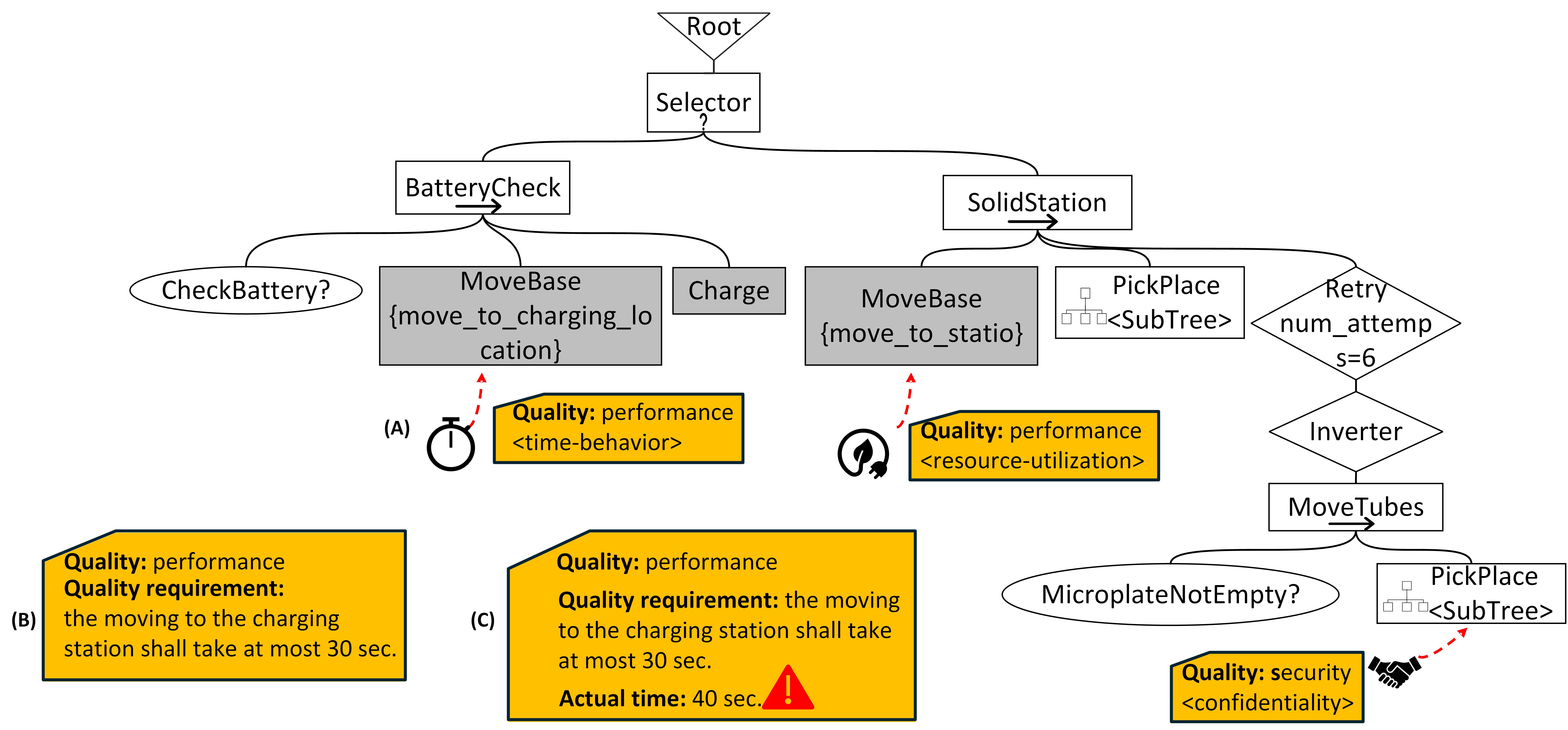}
	\end{center}
	\vspace{-5mm}
	\caption[BT example with qualities]{\footnotesize An example of a concrete model for representing the different quality concerns in \bts. For  \lstinline!MoveBase!, we provide representation across different development stages: (A) early-stage design time, (B) late-stage design time, and (C) run-time.}%
	\label{fig:figure-3}
	
	\vspace{-6mm}
	
\end{figure*}

A \lstinline!Quality! is an aspect related to the overall quality of a robotic task, and/or skill. Performance is an example of a quality that matters for the mobile laboratory mission (see \cref{fig:figure-3} described in \cref{subsec:concretemodel}). The attribute \lstinline!type:string! refers to the quality name, i.e. performance, security, etc.  This can be used to represent qualities in early design, which are not yet quantified.

A \lstinline!QualityRequirement! 
captures and formalizes a quality of a robotic task, and/or skill. An example of a quality requirement for the mobile laboratory mission is ``the moving to a charging station shall take at most 30 sec.''. This requirement specification corresponds to the performance quality. The class has two attributes. The attribute \lstinline!ID:string! corresponds to a unique identifier to distinguish between the different requirements. The attribute \lstinline!description:string! refers to the detailed description of the quality requirement. Ideally, this description should specify a numerical value to provide precise and measurable criteria for evaluating and capturing a quantified quality requirement in later design.
 
In our meta-model classes, we propose that a behavior-tree node can sufficiently satisfy a quality (\lstinline!satisfices! relation), but should fulfill (satisfy) a quality requirement (\lstinline!satisfies! relation). To highlight, we differentiate between a node satisficing a quality, potentially useful in early design stages, and a node satisfying a quality requirement in later design stages. To allow flexibility during design time, we allow the definition of a quality requirement without it belonging to a specific quality ((\lstinline!0..1! cardinality).  Thus, one has the freedom to express qualities without quality requirements in the early phases of design, and then to refine these qualities to quality requirements in later design, or to skip the early stages and define quality requirements directly.

\begin{table}[t]
	\vspace{-5mm} 
	\caption{\footnotesize Nodes, qualities, and quality requirements of the mobile laboratory robot.}
	\footnotesize
      \centering
    \scalebox{0.8}{
	\begin{tabularx}{\linewidth}{
			>{\small}p{30mm}
			>{\small}p{30mm}
			>{\small}p{5mm}
			>{\small}X}
		\toprule
		
		\textbf{Relevant}\\ \textbf{node/subtree} &  \textbf{Quality} & \textbf{ID}  &\textbf{Quality requirement} \\
		\midrule
		MoveBase \newline(BatteryCheck task) &  performance \newline <time-behavior> & rq1 & the moving to charging station shall take at most 30 sec (hard-constrain). \\ 
		MoveBase \newline(SolidStation task) & performance \newline<resource-utilization>  &  rq2 &after moving to the solid station, the robot should have at least a battery capacity of 3\% left. \\
		PickPlace \newline(MoveTubes task) & security \newline <confidentiality>  & rq3 &the information on the tube label shall be processed locally on the robot.\\   
		\bottomrule
		
		\label{tab:requirements}
	\end{tabularx}
    }
	\vspace{-10mm}
    
\end{table}

\vspace{-0.3cm}
\subsection{Example Instantiation of the Meta-Model}
\vspace{-0.2cm}
\label{subsec:concretemodel}

To demonstrate the instantiation of the proposed meta-model, we are using the behavior tree of the mobile-laboratory robot mission, see \cref{fig:figure-1}, and assuming the nodes need to adhere to important qualities.
We focus on qualities from the ISO/IEC 25010 product quality model~\cite{ISO25010}.
To the best of our knowledge, there is no quality standard for robotic missions. In March 2024, we searched for ISO standards for robotics using the official ISO website \url{https://www.iso.org/}, keyword robot and under the ISO/TC 299
robotics' committee. It returned 25  standards. The majority of them are concerned with safety, modularity, or performance. None provides a 
 comprehensive standard outlining quality requirements for robotic missions. We chose ISO/IEC 25010 to demonstrate the applicability of software quality standards in robotics.

For our demonstration, we chose from the ISO/IEC 25010 the following qualities: performance efficiency, specifically time-behavior and resource utilization, and security, specifically confidentiality.
\Cref{tab:requirements} presents the nodes/subtrees of the behavior-tree example and the qualities and quality requirements that they need to satisfice and satisfy, respectively. \Cref{fig:figure-3} illustrates an envisioned concrete model of \bts with the defined quality concerns for the mobile laboratory example. Displaying the quality concerns in the behavior tree in this way offers a concise overview of mission quality concerns, enhancing communication within the development team and facilitating ongoing monitoring to ensure high standards are maintained throughout robotics missions. 

\textbf{Use during early design-time.} In the early-stage design time (A in
\cref{fig:figure-3}), practitioners can start by identifying important qualities for their task without the need for an in-depth specification. They can flag the tree nodes that are relevant for satisficing them. In the example,
performance is flagged as a relevant quality for the \lstinline!MoveBase! skill.

\textbf{Use during late design-time.} In the late-stage design time (B in
\cref{fig:figure-3}), as things become clearer, quality requirements are added
to the behavior-tree model. This provides clarity regarding the different
components of the mission that need to satisfy the quality
requirements. In the example, a concrete and measurable performance
requirement is specified for the \lstinline!MoveBase! skill.

\textbf{Use at run-time.} At run-time (C in \cref{fig:figure-3}) monitoring quality requirements satisfaction can be accomplished by acquiring runtime
data. In the example, the performance of the  \lstinline!MoveBase! skill at
run-time is measured and can be tracked by humans working with
the system in case of violation. Note that we envision not only the current value to be displayed (actual time in C-\cref{fig:figure-3})
but potentially further information that might be beneficial about
the historical development of a measure over time (historical data in C-\cref{fig:figure-3}). If the mission is
executed multiple times, it is interesting to see if the actual time
only once deviates from the quality requirement or if that deviation
occurs frequently.


\vspace{-0.5cm}
\section{DSL and Implementation of the Behavior-Tree Extension}
\vspace{-0.3cm}
\label{sec:demonstration}

\begin{figure*}[t]
	\begin{center}
		\includegraphics[
		width=0.9\linewidth
		]{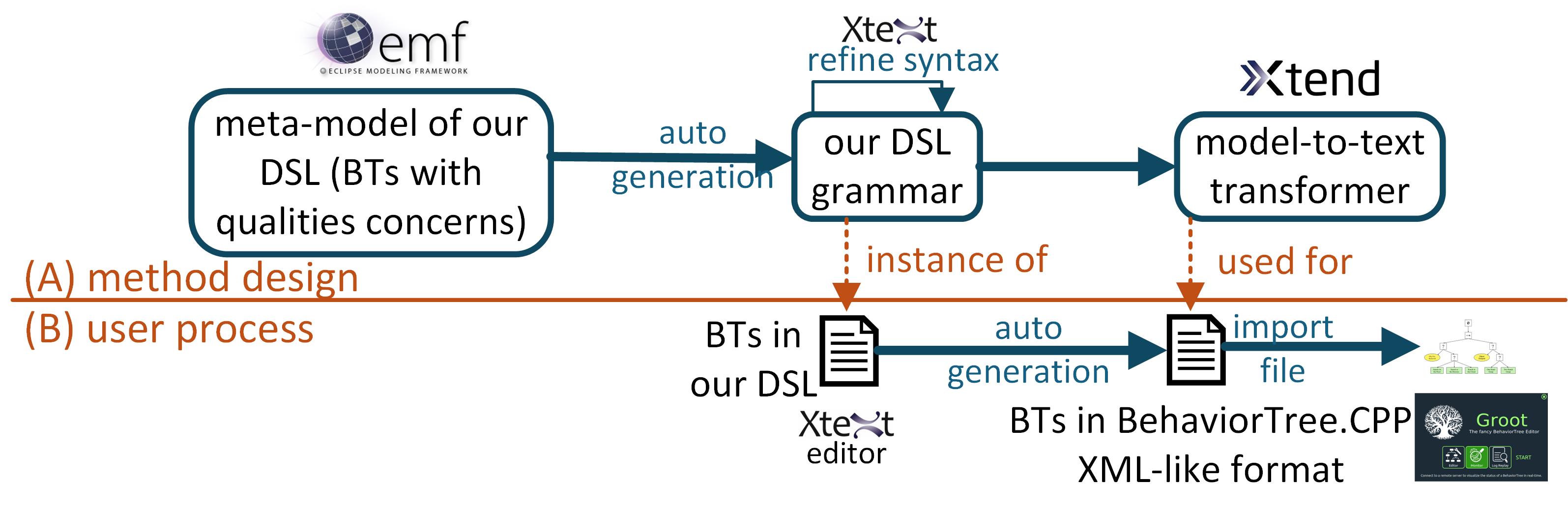}
	\end{center}
	\vspace{-5mm}
	\caption[MDE used method]{\footnotesize An overview of (A) our design process of our DSL using MD practices and (B) the user process to use the DSL.}%
	\label{fig:figure-6}
	
	\vspace{-6mm}
	
\end{figure*}

We designed our model extension to be generally applicable to all kinds of quality requirements and provide flexibility to domain experts in the different phases of developing robotics missions. To showcase the applicability of our approach and support domain experts, we designed a domain-specific language (DSL).  We want our DSL to be integrated into existing robotics frameworks to leverage existing functionalities. 
We chose to build a textual DSL as an extension of \BTCPP (see \cref{sec:background}) with the functionality to define qualities and quality requirements.  
We chose \BTCPP since it employs good software engineering practices in its design, and it has a GUI, Groot, that can be used during the design time and execution time of a robotic mission.

To create our DSL, we employ model-driven (MD) practices. \Cref{fig:figure-6} shows an overview of the used MD practices from the perspective of both the method designer (us) (indicated as \textit{(A) method design}) and the perspective and the user of our DSL, highlighted as \textit{(B) user process} in the bottom part of the figure. 
We used the Eclipse Modeling Framework (\url{https://eclipse.dev/modeling/emf/}, EMF) to create a meta-model similar to \Cref{fig:figure-2} with more types of execution nodes, decorator nodes, and control nodes taken from the range of supported types by \BTCPP. We chose Eclipse because it offers a comprehensive and powerful environment for MD engineering and DSL development \cite{steinberg2008emf}. The resulting meta-model is too large to show in the paper, so we provide a high-resolution image of the detailed meta-model in our online appendix \cite{appendix:online}. Our DSL meta-model can be seen as a meta-model for \BTCPP with our proposed extension of quality concerns. 

To create a textual DSL conforming to our meta-model, we use Xtext \cite{efftinge2006oaw}, a textual modeling framework (TMF) for Eclipse. We use Xtext since it auto-generates a grammar from a created meta-model, including the automatic generation of editors, parsers, and other tooling. We edited the auto-generated grammar to design a simple and initiative syntax for creating \bts. 

\Cref{ls:xtextnotation} shows an excerpt from our DSL grammar allowing one to define a behavior-tree node, sequence node, quality and quality requirement. For a sequence node, it is optional to assign a node both qualities and quality requirements. We want to highlight that we allow the flexibility to specify a quality requirement even if its corresponding quality is unknown (line 7 by using \lstinline!satisfies!). This DSL design provides flexibility and support in different stages of mission design. It is possible and optional to assign a sequence node an ID, a name, or other needed parameters. We opt out in the EBNF to show similar details for simplicity and focusing only on the quality part. Indentation is used to indicate the definition of a new child of a node (line 7 \lstinline!indent! in purple). Similar to the Sequence node, other node types can be defined in our DSL following the same style.
For the rest of the node types, we refer the reader to Xtext grammar in our online appendix.

We aim to keep the control-nodes notation simple and similar to the literature, as well inspired by \BTCPP.
For sequence, selector and parallel, we use the same symbol notation as in \cref{tab:btsyntax} (line 7 in \cref{ls:xtextnotation} \lstinline!->!). For all the other types, such as decorator nodes and action nodes, we use the same naming schema as in \BTCPP to reduce the learning curve (we refer the reader to \BTCPP documentation).
For more details about our supported types, we refer the reader to our online appendix \cite{appendix:online}.
\vspace{-1mm}
\begin{lstlisting}[style= xtextstyle, caption={\label{ls:xtextnotation} The EBNF grammar notation for an excerpt of our DSL with focus on defining quality and quality requirements. Our DSL syntax and keywords are in blue.}]	
	BehaviorTree = "BehaviorTree" "ID" "=" <STRING> indent TreeNode
	TreeNode = (LeafNode | ControlNode | DecoratorNode | SubTree)
	ControlNode = (SequenceNode | ParallelNode | ...)
	DecoratorNode = (RepeatNode | InverterNode | ...)
	LeafNode = (ActionNodeBase | ConditionNode)
    
	SequenceNode = "->" ["("  ["satisfices" Quality+] ["satisfies" QualityRequirement+]
                      ")"] indent TreeNode+
    
	Quality = "Quality" "=" <STRING>  {"(" QualityRequirement+")"}
    
	QualityRequirement = "QualityReq" "ID" "=" <STRING> 
                           "description" "=" <STRING>
    
  STRING = {a-zA-Z0-9}
  indent = \tab	
\end{lstlisting}
\vspace{-2mm}

By leveraging the support in Eclipse for model-to-text transformation using Xtend \cite{bettini2016implementing}, we created an Xtend generator to transform automatically any model created with our DSL to a \BTCPP XML-like code format.  
The automatic transformation allows an easier integration of our DSL to existing workflow in robotics projects. 

\BTCPP does not have a notation for qualities and quality requirements in its XML-like language. To have the quality specifications explicit, we mapped the specifications in our DSL to the description part of a node definition in \BTCPP.  If a quality requirement is a hard constraint that leads to a node failure or success, the user of our DSL should use the keywords \lstinline!FailureIf! and \lstinline!SuccessIf! in the description followed by the requirement. The XML-like language of \BTCPP allows the specification of the conditions that capture when a node fails and/or succeeds. Therefore, we express hard-constraint quality requirements by setting the corresponding failureIf/successIf fields in the \BTCPP code. 
All the materials described here, like the generator code, are provided in our online appendix \cite{appendix:online}.

\begin{figure*}[t]
	\begin{center}
		\includegraphics[
		width=\linewidth
		]{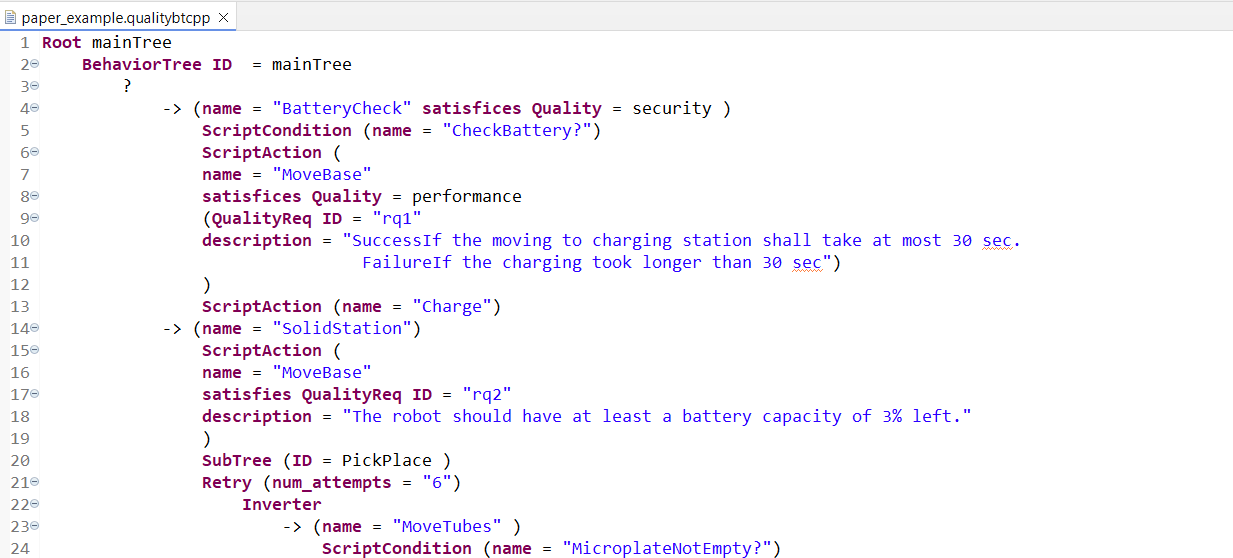}
	\end{center}
		\vspace{-5mm}
	\footnotesize\caption[Example of our DSL syntax]{Our DSL representing part of the laboratory mission in Xtext editor.}%
	\label{fig:figure-4}
	
		\vspace{-5mm}
	
\end{figure*}

\begin{figure*}[t]
	\begin{center}
		\includegraphics[
		width=0.9\linewidth
		]{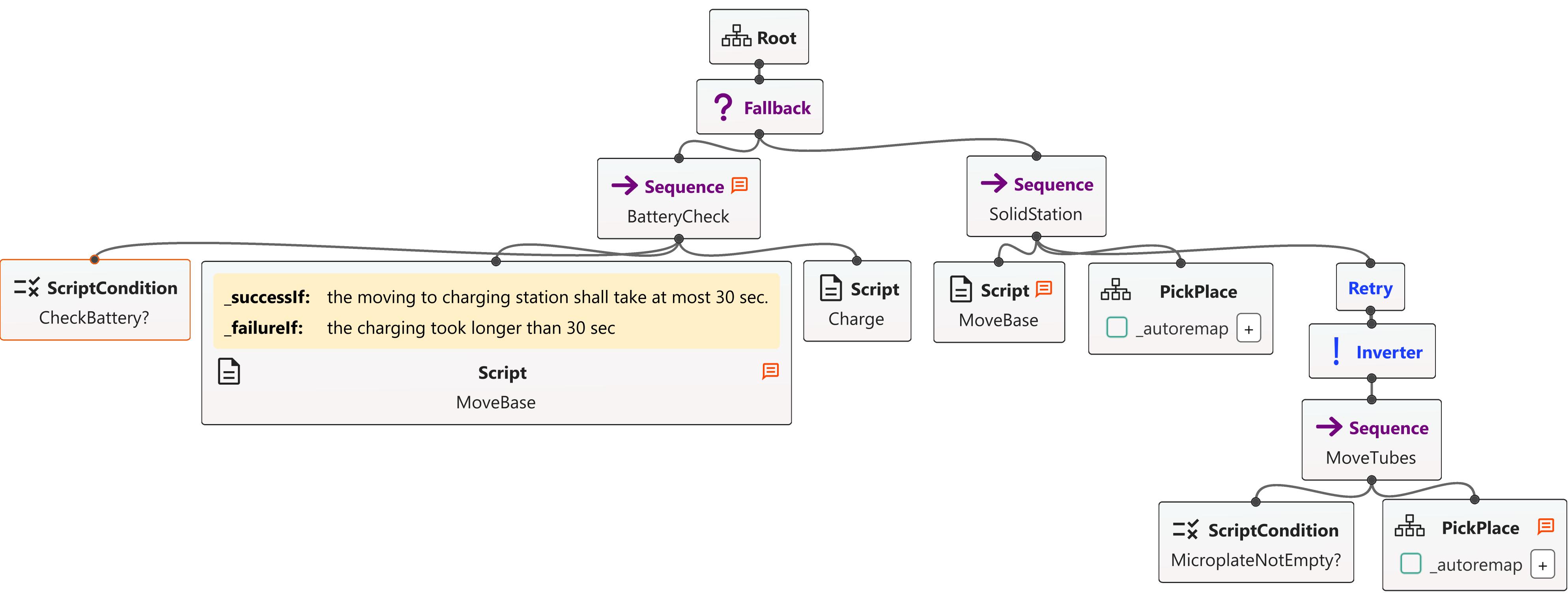}
	\end{center}
	\vspace{-5mm}
	\footnotesize\caption[Groot view of the transformation]{The view of laboratory mission in Groot after importing the auto-generated code for \BTCPP from our DSL. }%
	\label{fig:figure-5}
	
	\vspace{-6mm}
	
\end{figure*}

To demonstrate our DSL, we are taking the mobile-laboratory robot mission,  defining important qualities (see \cref{tab:requirements}), and then showcasing the usage of our DSL to define them in \bts. 
\cref{fig:figure-3} shows an overview of the important qualities in this mission represented in the behavior-tree model. \Cref{fig:figure-4} represents part of the laboratory mission in our textual-DSL.

\textbf{Use during early design-time} In \cref{fig:figure-4} (line 6-8), performance is flagged as a relevant quality for the \lstinline!MoveBase!.
By capturing this information at the early-stage design time, \lstinline!MoveBase! could be automatically flagged as a concern for designers of a mission. The same applies to the other qualities represented.

\textbf{Use during late design-time} In \cref{fig:figure-4} (line 9-12), a concrete and measurable performance requirement is specified for the \lstinline!MoveBase!. 
Using the keywords \lstinline!SuccessIf! and \lstinline!FailureIf! indicates that a quality requirement is a hard constraint and leads to the success or failure of a node. 

\Cref{fig:figure-5} shows the mission in Groot after importing the auto-generated \BTCPP code from the created model in our DSL. 

\textbf{Use at run-time} Once the mission specifications are written in our DSL in the Xtext editor, the user receives an auto-generated file with code in \BTCPP XML-like language. \Cref{fig:figure-5} shows the mission in Groot after importing the auto-generated file from the created model in our DSL. The nodes annotated with the comment bubble symbol represent that there is information in the descriptor part of the nodes, which can be displayed after clicking the node. The yellow part of the \lstinline!MoveBase! node highlights our mapping of the hard-constraint performance requirement into the failure and success part of the node. 
At run-time, monitoring quality requirements satisfaction can be accomplished by acquiring run-time data. 
In the example, the performance of the \lstinline!MoveBase! skill at run-time is measured and can be tracked by humans working with the system in Groot. Currently,  monitoring of quality satisfaction is done manually by the developers. We envision improving this aspect by doing automatic monitoring in future work. 
The current way Groot displays our mapped information about the qualities and quality requirements is not optimal and does not provide a clear overview. We believe using our DSL in combination with \BTCPP should provide a better understanding of the displayed \bt. We think there is an opportunity for the software community in robotics to improve the existing behavior-tree frameworks accommodating the view of qualitites.


\vspace{-0.5cm}
\section{Preliminary Evaluation of the Behavior-Tree DSL}
\vspace{-0.3cm}
\label{sec:evaluation}

We conducted a preliminary evaluation on the feasibility of using the developed DSL to specify quality concerns of robotic missions. We also aimed to evaluate the benefits of using the DSL and assess further needs for such a DSL.

\textbf{Study Design.} The evaluation was guided by two research questions:

\noindent \textbf{RQ1:} How useful is the designed DSL to model quality concerns in behavior trees in practice?

\noindent\textbf{RQ2:} What would prevent people from using the designed DSL?

With RQ1, we aim to assess the extent to which our tool provides support for specifying quality concerns in \bts.
With RQ2, we want to assess the needs and features that are missing in the designed DSL and the drawbacks of our DSL. 
By answering those questions, we provide preliminary empirical data about the feasibility of our DSL and its usefulness in practice.
 
To demonstrate our DSL, we used the mobile laboratory robotic mission and the quality concerns in \cref{tab:requirements}. We held individual sessions, 20-30 minutes long, with different practitioners and researchers. We picked individuals with prior knowledge of behavior-tree models. In the session, one of the authors presented a behavior-tree model for the mobile laboratory robotic mission using high-level abstract notation (see \cref{fig:figure-1}). This was followed by presenting the DSL and the way to create the mission using it. Then, we introduced the qualities and quality requirements and showed different ways to specify them using our DSL, corresponding to the different stages of design: early-design time where only qualities might be known, and late-design time where quality requirements are specified where they might belong to a quality or not. For quality requirements, we presented the two cases when the requirements are soft constraints and hard constraints and how to specify them in our DSL. Then, we presented the integration of our DSL with \BTCPP by showing the resulting XML-like code and importing it into Groot (see \cref{fig:figure-5}). Finally, we shared a survey with questions about the participants' backgrounds and asked them to evaluate our DSL. \Cref{tab:survey} presents the non-demographic part of the questions in the survey and the research questions they answer. The survey collected a mix of qualitative and quantitative data. We used thematic analysis \cite{guest2011applied} to identify and analyze patterns in the provided answers.

\begin{table}[t]
	\vspace{-5mm} 
	\caption{Questions from the survey and their relations to the RQs.}
	  \centering
    \scalebox{0.8}{
	\begin{tabularx}{\linewidth}{
			>{\small}p{15mm}
			>{\small}p{85mm}
			>{\small}p{30mm}}
		\toprule
		
		\textbf{Research question}  &  \centerfirst{\textbf{Survey question}} & \textbf{Data}  \textbf{type}\\
		\midrule
		& Have you used behavior tree models? & Nominal data\\
		& Have you used  BehaviorTree.CPP for creating a behavior tree? & Nominal \newline data\\
		RQ1 & Do you consider qualities and quality requirements in your robotic missions? & Open-ended \newline responses\\
		& I think that I would like to use the presented DSL in my own work.&  Scale data \newline (1-5)\\
		& I think the benefits of using the presented DSL in my work are: & Open-ended \newline responses
		\\
		\midrule
		RQ2 &   I think what is missing in the presented DSL to use it in my work is: & Open-ended \newline responses
		\\
		
		\bottomrule
		\label{tab:survey}
	\end{tabularx}%
    }
	\vspace{-10mm}
\end{table}

\textbf{Results.}
We ran the evaluation with six participants (half from industry and half from academia). All of them were familiar with the basic node types of \bts in robotics and 4/6 used \BTCPP before. It was expressed that different qualities mattered to them, such as safety, performance and reliability. Two participants, a practitioner and an academic, did not consider qualities in their work, but they saw the value of doing that as early as possible. 

As an answer to the likelihood of using our DSL in their work, all participants expressed it would be feasible to use it.
All participants saw the benefit of using our DSL to express quality concerns in the early stages of the development of robotic missions and directly in \bts. They appreciated that the DSL provided an overview of important qualities directly in \bts. One participant saw the DSL output as notes to the developers to help them shape the implementation of execution nodes. Two participants stated that the syntax of our DSL is light and initiative, especially as we are using a similar notation from the literature for control nodes. 

Participants stated three desired features and two drawbacks for future iterations.
The first requested feature was the need for run-time monitoring of the qualities' compliance. The second feature was providing the probability of a quality compliance e.g., 90\% chance that the robot reaches the charging station in 30 sec. 
The final requested feature for our DSL was to easily specify and reuse cross-cutting quality concerns. Currently, it is only possible by using the same identifier for cross-cutting requirements and then having different nodes contributing to the same requirement. For all mentioned features, we see potential to integrate those needs in future work and for developers to consider them when working with qualities in \bts. 

In terms of the drawbacks, one is concerned with the technology used to develop the DSL.  One practitioner expressed that if the company does not use Eclipse, then it might be better to develop the DSL as a standalone Python library. We chose Eclipse to leverage the support of model-driven (MD) approaches when creating our DSL. To the best of our knowledge, there is only one library in Python, textX (\url{https://github.com/textX/textX}), with minor support for MD \cite{dejanovic2017textx}. At the beginning of our work, we tried using that library;  however, we found that it fell short in the provided support compared to Eclipse, and was cumbersome to use. Finally, a participant expressed that a graphical DSL might be better to use at the early stages, compared to our textual DSL. We think the integration with \BTCPP and using its GUI could overcome such a problem. We believe that further investigations are needed for both points in our future work. The results of the survey are available in our online appendix. 

\textbf{Study Limitation.}
The results of our preliminary evaluation of the proposed behavior-tree DSL show the need for similar tools to express and consider quality concerns from the early stages of the robotics missions in behavior trees. However, the small size of the study sample threatens the generalizability of the results.  Although the results are not generalizable, they provide a first step towards understanding the feasibility of similar solutions. We plan to mitigate this threat in future studies and expand the sample size. 

Another threat that could affect our results is the scalability of the proposed method and DSL.
We selected an example inspired by the robotics company Kuka for the automation of a laboratory \cite{burger2020mobile} with a mission size and a number of quality concerns that allowed us to illustrate the approach comprehensibly. We acknowledge that the evaluation with a limited mission size and number of quality concerns will need to be extended when we aim to focus on larger, real-world scenarios. In general, the scalability of DSLs and models is a known shortcoming in the model-driven engineering community and there has been ongoing research to overcome it \cite{bucchiarone2020grand}. 
We plan to explore and evaluate the scalability aspect of the proposed approach in future studies and build on top of other research in this area.

\vspace{-0.5cm}
\section{Related Work}
\vspace{-0.3cm}
\label{sec:relatedWork}

In the last decade, behavior trees have been used for modelling the non-player characters in computer games and modelling robot behavior \cite{Mcquillan2015,colledanchise2018behavior,iovino2022survey}. Existing work in robotics has provided a framework to unify the syntax and semantics of the behavior-tree model \cite{marzinotto2014towards}. Rovida et al.~\cite{rovida2017extended} extended the notation of behavior trees to add pre- and post-conditions to the action nodes. Others worked on improving the already existing node types of the model \cite{colledanchise2019analysis,giunchiglia2019conditional} or making behavior trees state-aware \cite{de2022state}. The former work focuses on improving the execution aspect of behavior trees. Our work focuses on representing other concerns of the robotics mission, specifically the qualities and quality requirements of the missions, which none of the previous work did.   

Considering qualities and requirements in robotic systems is not a novel concept. Steck et al.~\cite{steck2010towards} proposed a model-driven development process that incorporates non-functional properties and quality of service parameters of the robotic system, aiming for resource-aware utilization. Non-functional properties and quality of service parameters of robotics systems provide the basis for defining quality requirements. Reichardt et al. \cite{reichardt2013software} have proposed a design framework to support certain qualities in robotic systems, while \cite{monthe2016rsaml} focused on providing a domain-specific modelling language for describing the robotic system architecture that captures real-time requirements.  
The previous work focused mainly on the overall robotic system or the system architecture rather than the robotic missions. Also, none of the previous works has focused on behavior trees as a behavior model for mission specification.
Our work focuses on capturing robotic missions-related qualities in behavior trees to have a better overview of the mission's qualities and facilitate better monitoring of their satisfaction.

\vspace{-0.5cm}
\section{Conclusion and Future Work}
\vspace{-0.3cm}
\label{sec:conclusion}

In this work, we provide a first step to enhance \bts by introducing a way to capture quality concerns. 
We provided a meta-model extension for behavior trees and demonstrated the applicability of our model by creating a DSL to support the extension. Our DSL and metamodel are connected to popular robotic tooling, allowing for direct and easy access by roboticists.  We conducted a preliminary evaluation of the DSL, where participants expressed the need for such a tool specifying quality concerns in \bts. However, further development of our approach is needed to integrate run-time monitoring of the requirements compliance and the definition of cross-cutting quality concerns in nodes.

In the future, we would like to expand the evaluation with a larger sample of practitioners and a hands-on usability study. 
Getting insights about the applicability and advantages/disadvantages of the proposed model and DSL would be one of our goals in the evaluation study. We would also like to check if other components in the model are needed and if different models are required for different qualities in the industry. Finally, our DSL is only one way to implement a \bts quality extension, so in future evaluation studies we want to check the usability of our approach and if other forms of DSLs or graphical representations are needed.

\noindent\textbf{Acknowledgments.}
We thank Ricardo Caldas for part of the model-to-text transformer from an earlier unpublished work with the first author. 
This work was supported by Wallenberg AI, Autonomous
Systems and Software Program (WASP), Knut and Alice Wallenberg Foundation.

\vspace{-0.5cm}

\bibliographystyle{splncs03_unsrt.bst}
\bibliography{main.bib}

\end{document}